\documentclass[preprint2]{aastex63}

\usepackage{graphicx}
\usepackage[suffix=]{epstopdf}
\usepackage{natbib}
\usepackage{amsmath}
\usepackage{url}
\usepackage{xspace}
\usepackage{mathrsfs}

\newcommand{\xblue}{192,489\xspace} 
\newcommand{\xred}{138,823\xspace} 
\newcommand{\xpink}{25,817\xspace} 
\newcommand{\xblack}{134\xspace} 

\begin{document}

%%%%%%%%%%%%%%%%%%%%%%
\title{Searching the SETI Ellipsoid with Gaia}

\correspondingauthor{James. R. A. Davenport}
\email{jrad@uw.edu}

\author[0000-0002-0637-835X]{James R. A. Davenport}
\affiliation{Astronomy Department, University of Washington, Box 951580, Seattle, WA 98195, USA}

\author{B\'arbara Cabrales}
\affiliation{Department of Astronomy, Smith College, Northampton, MA 01063, USA}

\author[0000-0001-7057-4999]{Sofia Sheikh}
\affiliation{Department of Astronomy, University of California Berkeley, Berkeley CA 94720, USA}
\affiliation{SETI Institute, 339 N Bernardo Ave Suite 200, Mountain View, CA 94043}

\author[0000-0003-4823-129X]{Steve Croft}
\affiliation{Department of Astronomy, University of California Berkeley, Berkeley CA 94720, USA}

\author{Andrew P. V. Siemion}
\affiliation{Department of Astronomy, University of California Berkeley, Berkeley CA 94720, USA}
\affiliation{SETI Institute, 339 N Bernardo Ave Suite 200, Mountain View, CA 94043}

\author{Daniel Giles}
\affiliation{SETI Institute, 339 N Bernardo Ave Suite 200, Mountain View, CA 94043}

\author{Ann Marie Cody}
\affiliation{SETI Institute, 339 N Bernardo Ave Suite 200, Mountain View, CA 94043}

\shorttitle{The SETI Ellipsoid with Gaia}
\shortauthors{Davenport et al.}

%%%%%%%%%%%%%%%%%%%%%%%%%%%%%%
\begin{abstract}
The SETI Ellipsoid is a geometric method for prioritizing technosignature observations based on the strategy of receiving signals synchronized to conspicuous astronomical events. 
Precise distances to nearby stars from Gaia makes constraining Ellipsoid crossing times possible.
Here we explore the utility of using the Gaia Catalog of Nearby Stars to select targets on the SN 1987A SETI Ellipsoid, as well the Ellipsoids defined by 278 classical novae. 
Less than 8\% of stars within the 100 pc sample are inside the SN 1987A SETI Ellipsoid, meaning the vast majority of nearby stars are still viable targets for monitoring over time. 
We find an average of 734 stars per year within the 100 pc volume will intersect the Ellipsoid from SN 1987A, with $\sim$10\% of those having distance uncertainties from Gaia better than 0.1 lyr. 
\end{abstract}

%%%%%%%%%%%%%%%%%%%%%%%%%%%%%%
\section{Introduction}

In developing robust searches for technosignatures -- signs of technological activity from extraterrestrial sources -- we must confront choices on both the types of signals that will be explored from our data, as well as which stars to observe and when \citep{tarter2001}. While this is broadly true for {\it all} studies in observational astronomy that explore detection limits on rare phenomena, it is especially so in the search for extraterrestrial intelligence (SETI). Here we focus on the latter challenge, identifying which nearby stars to monitor and when based on their locations relative to a coordinating beacon or event.

Transmitting signals in coordination with conspicuous galactic-scale events has been identified as an efficient means for generating simple interstellar beacons \citep[e.g.][]{corbet1999}, and possibly even in establishing two-way communications \citep{seto2019}. These events must be sufficiently rare and noteworthy (e.g. nearby supernovae), such that extraterrestrial astronomers could reasonably assume that unknown observers (i.e. us) would take notice. The geometric approach to identifying stars that could be generating such synchronized signals at a given point in time is known as the ``SETI Ellipsoid'' \citep{lemarchand1994}. The limiting factor for selecting targets based on this approach has traditionally been a lack of precise distances for nearby stars \citep{makovetskii1977}.

Gaia provides a revolutionary step forward in our understanding of the true locations of stars in the Milky Way \citep{gaia}. While the Hipparcos mission estimated distances for 118k stars in the solar neighborhood \citep{perryman1997}, the latest release from Gaia provides parallaxes for over 1.5 billion sources \citep{gaia_edr3}. 
With more than $\sim$200x better astrometric precision than Hipparcos, Gaia enables $<10\%$ distance uncertainties for for stars out to several kpc \citep{bailer-jones2021}.
This remarkable distance precision directly translates into lower uncertainties on the timing for signal coordination along the SETI Ellipsoid.

In this paper we demonstrate the utility of Gaia in defining targets for monitoring along the SETI Ellipsoid. In \S\ref{sec:elip} we review the geometry used in the SETI Ellipsoid framework. We focus our discussion in \S\ref{sec:gncs} around the sample of 331k stars from Gaia within 100 pc of the Sun, and in \S\ref{sec:targets} we demonstrate how to select targets of interest from this sample over time in coordination with SN 1987A. 
We provide an overview of target yields from other possible coordinating events in \S\ref{sec:future}. Finally, in \S\ref{sec:discuss} we conclude with a summary and discussion of future work.

%%%%%%%%%%%%%%%%%%%%%%%%%%%%%%
\section{The SETI Ellipsoid}
\label{sec:elip}

\begin{figure*}[!ht]
\centering
\includegraphics[width=6in]{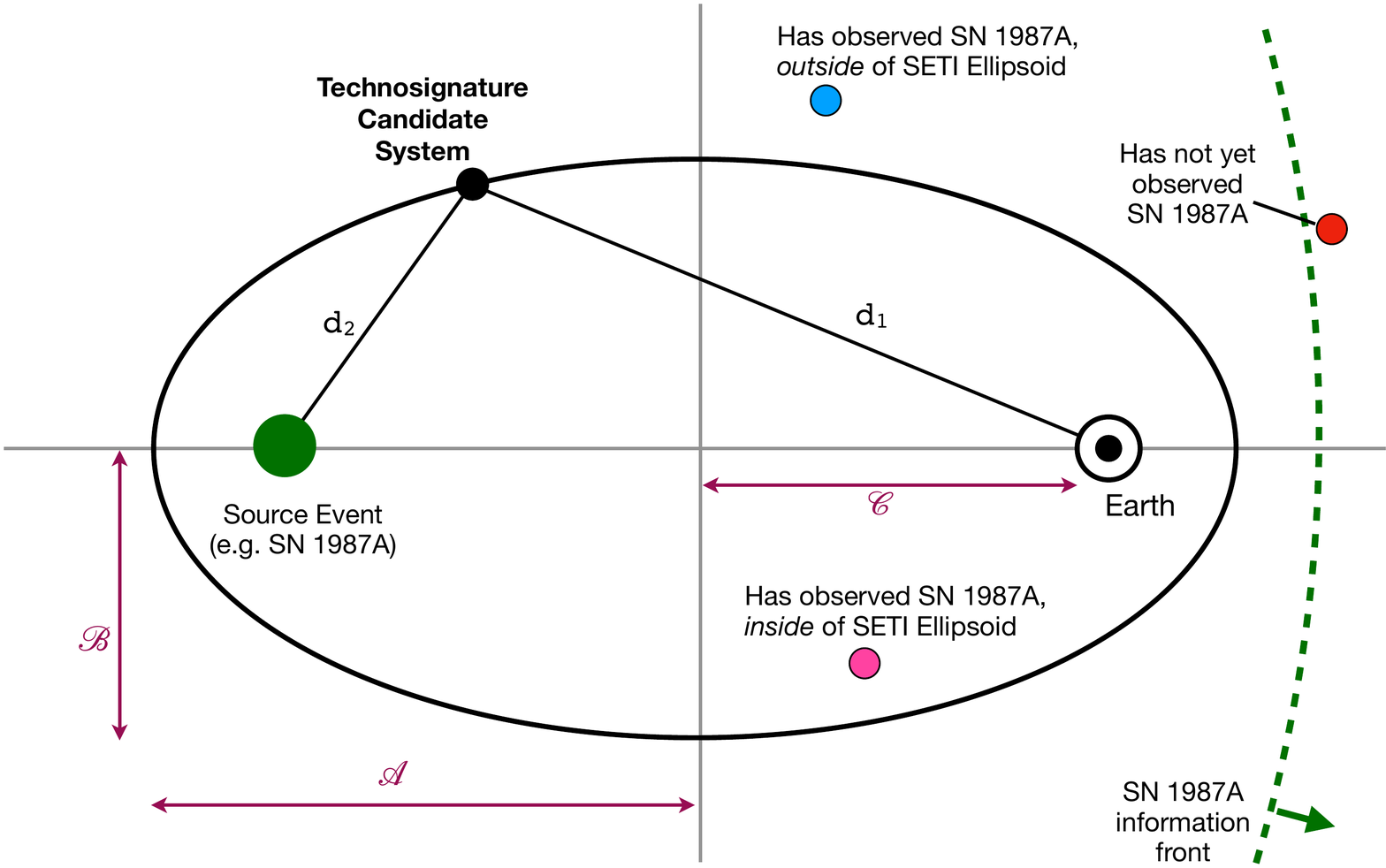}
\caption{
Schematic diagram of the SETI Ellipsoid framework. A civilization (black dot) could synchronize a technosignature beacon with a noteworthy source event (green dot). The arrival time of these coordinated signals is defined by the time-evolving ellipsoid, whose foci are Earth and the source event. Stars outside the Ellipsoid may have transmitted signals in coordination with their observation of the source event, but we those signals have not reached Earth yet (blue dot).
For stars far {\it inside} the Ellipsoid (pink dot), we have missed the opportunity to receive such coordinated signals.
}
\label{fig:elip}
\end{figure*}

The ``SETI Ellipsoid'' is a geometric framework for identifying signals (i.e. beacons) that are in synchronization or coordination with noteworthy astronomical phenomena \citep{tang1976,lemarchand1994}. Under this scheme, shown in Figure \ref{fig:elip}, an extraterrestrial agent would observe a rare event (e.g. a nearby supernova), and shortly thereafter broadcast a conspicuous signal indicating they have observed the event. Utilizing these rare events allows extraterrestrial agents to more efficiently operate  beacons, as they need to broadcast signals only at specific times. 
The source event provides a natural focal point of attention for other astronomers (i.e. us), acting as ``Schelling Points'' to facilitate communication between unknown observers \citep{wright2017}.

%%%%%

This synchronized communication approach has been suggested as a basis for technosignature searches previously \citep[e.g.][]{makovetskii1977,lemarchand1994}. We also note that the basic geometry of the SETI Ellipsoid is identical to the study of light echos around supernovae \citep[e.g.][]{chevalier1988}.
Here we provide a brief review of the framework.

The SETI Ellipsoid is so named because it leverages the geometric properties of ellipses. As shown in Figure \ref{fig:elip}, we consider an ellipsoid whose foci are the location of the source event (e.g. a nearby supernova) and the Earth. 
The ellipsoid expands in time, and stars that intersect the surface of the ellipsoid could choose to produce a signal coordinated to leave near-synchronously with the arrival of the event. 
The time that synchronized signals would arrive to us from a given star is based on the distance to the star ($d_1$), and the distance from that star to the source event ($d_2$), both in units of lyr. Following the discussion by \citet{lemarchand1994}, the Ellipsoid at any given time can be defined by:

\begin{equation}
    d_1 + d_2 = 2\, \mathscr{A} = 2\, \mathscr{C} + T
\label{eqn}
\end{equation}

\noindent
where, as shown in Figure \ref{fig:elip}, $\mathscr{A}$ is the semi-major axis of the ellipsoid in lyr, $\mathscr{C}$ is the foci distance from the center of the ellipsoid in lyr ($2\, \mathscr{C}$ is the distance from Earth to the synchronizing event), and $T = c\, t$ is the elapsed time since the synchronizing event was observed on Earth ($t$) times the speed of light ($c$), in units of lyr.

Stars can be grouped into four categories based on this framework: 
1) Stars that have not yet observed the source event and therefore cannot have transmitted a synchronized signal\footnote{Unless they are anticipating a noteworthy astrophysical event, as \citet{seto2019} suggest.}. 
2) Stars that have observed the source event, but if they have transmitted a synchronized signal it has not yet reached Earth.
3) Stars whose potential synchronized signal would be arriving at Earth now (i.e. those intersecting the Ellipsoid).
4) Stars for which any synchronized signal must have arrived at Earth in the past.

\begin{figure}[]
\centering
\includegraphics[width=3.25in]{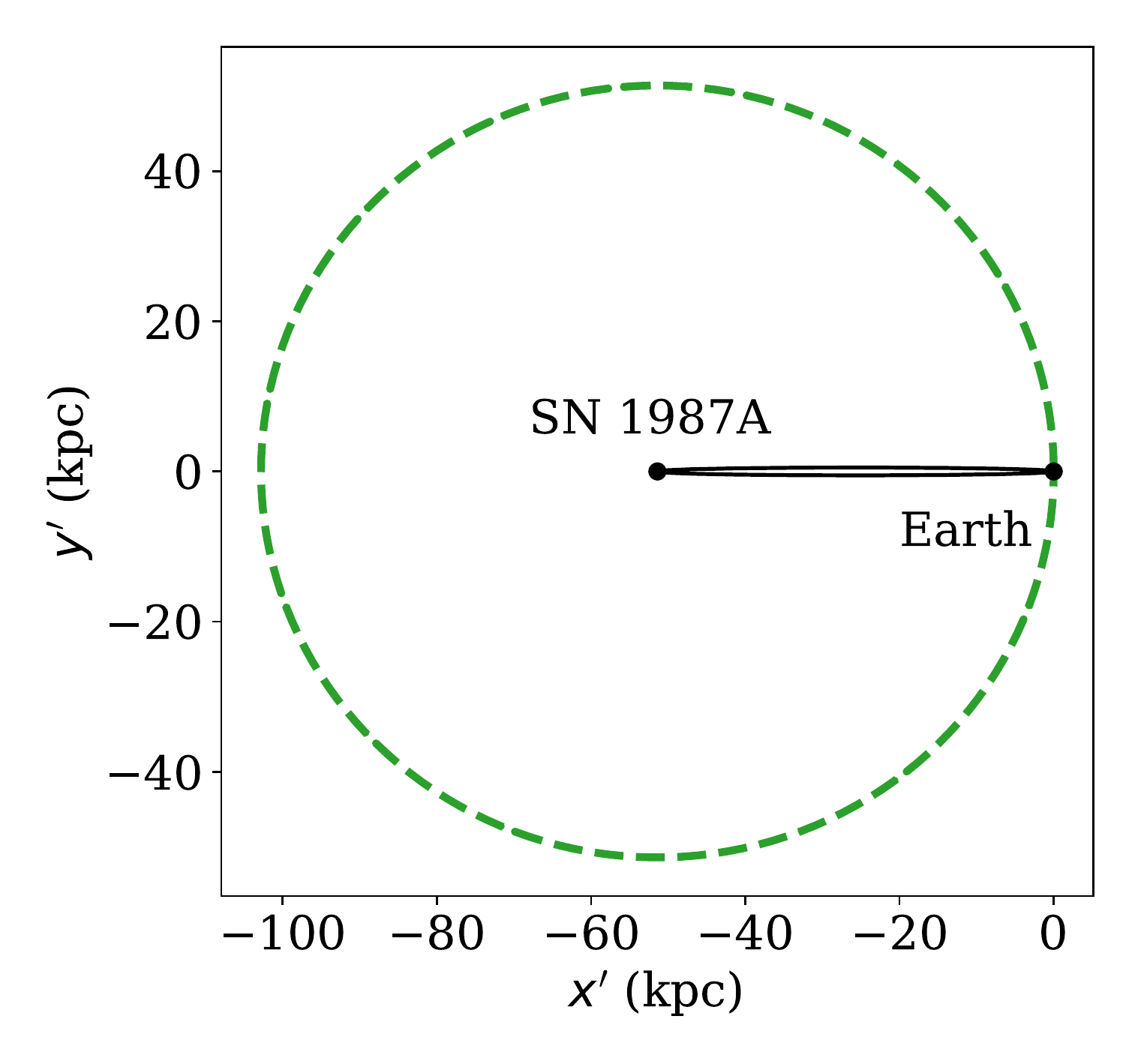}
\caption{
To-scale drawing of the information front (green dashed line) and the SETI Ellipsoid (black solid line) for SN 1987A, in an arbitrary coordinate frame oriented along the line of sight to the Large Magellanic Cloud. Though SN 1987A was observed more than 35 years ago, very few synchronized signals could have reached us thus far.
}
\label{fig:toscale}
\end{figure}

In Figure \ref{fig:elip}, and throughout most of this paper, we focus on SN 1987A as a noteworthy source event to consider in generating a SETI Ellipsoid. In Figure \ref{fig:toscale} we show a {\it to-scale} drawing of the SETI Ellipsoid and the information front (i.e. when the event is observed) from SN 1987A. We use the distance for SN 1987A of 51.4 $\pm$ 1.2 kpc, defined by \citet{panagia1999}.
While light from the supernova has reached a great many stars in the Milky Way, the Ellipsoid carves out a tremendously small volume of space still. Though SN 1987A was observed more than 35 years ago, the potential for exploring galactic-scale signal coordination with this event is just beginning.

%%%%%

The SETI Ellipsoid framework itself offers few clues as to the nature of any beacon or signal. Instead, it provides a straightforward method for selecting targets to observe in time. Given accurate distances, Equation 1 can be solved either to plan when a sample of stars should be observed to maximize the likelihood of receiving synchronized signals, or equivalently can determine which stars in a given set of observations (e.g. in an archive) intersect the Ellipsoid.
Since the synchronization is governed by light travel time, uncertainties in distances to stars, to the source event, and in the dates of source events (e.g. for historical supernovae) directly impact timing uncertainties for receiving signals, and must be propagated appropriately when solving Equation 1.

%%%%%%%%%%%%%%%%%%%%%%%%%%%%%%
\section{The Gaia Catalog of\\ Nearby Stars}
\label{sec:gncs}

The most recent data release from the Gaia mission, EDR3, provides the largest and most precise sample of astrometry ever gathered for stars in our Galaxy, containing more than 1.5 billion sources with parallax measurements \citep{gaia_edr3}. This latest catalog has parallax measurements that are 30\% more precise than the previous Gaia release. Using a model that accounts for both variable interstellar extinction and the Gaia magnitude limit, \citet{bailer-jones2021} have generated probabilistic distance estimates for 1.35 billion stars in Gaia EDR3. The increased precision from Gaia EDR3, and the probabilistic distances from \citet{bailer-jones2021}, enable us to make both accurate and precise estimates for when nearby stars will intersect the SETI Ellipsoid for a given source event.

The Gaia Catalog of Nearby Stars \citep[GCNS;][]{GCNS} contains a benchmark sample of 331k stars from the Gaia EDR3 release with robust distances within 100pc of the Sun. The GCNS provides probabilistic distances to each star, and is estimated to be $\sim$92\% complete for stars down to spectral type M8. All GCNS targets are also brighter than $G\sim20.5$ mag. This ensures we can explore a wide range of spectral types with modest aperture telescopes, and with a high confidence in our Ellipsoid timing. The GCNS is therefore ideal for selecting targets for monitoring with the SETI Ellipsoid.

%%%%%
\begin{figure}[]
\centering
\includegraphics[width=3.25in]{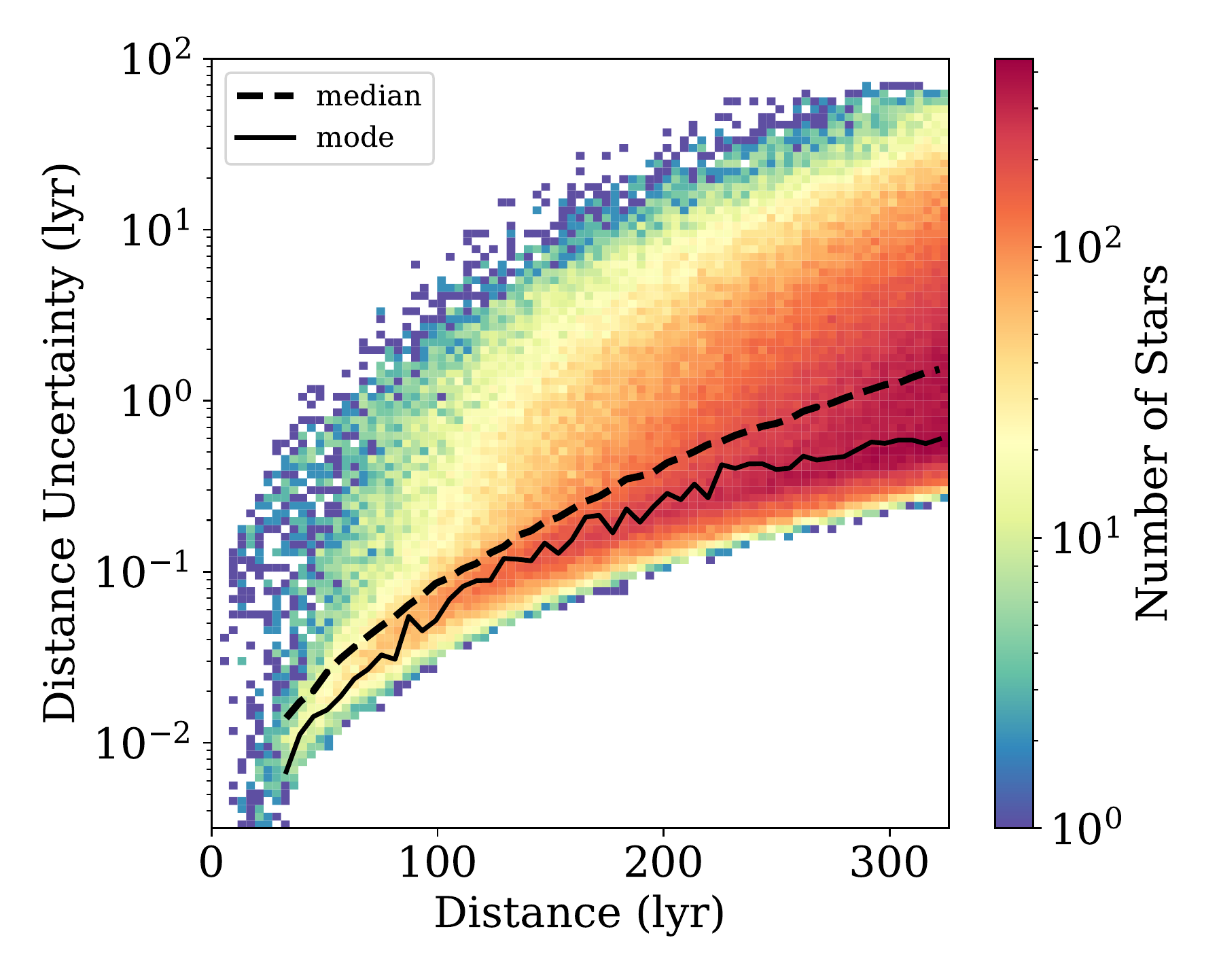}
\caption{
Two dimensional histogram of the symmetrized distance errors for 331k stars in the GCNS, with the median (dashed line) and mode (solid line) distance errors shown. Typical  uncertainties are less than 1 lyr for all stars within the 100 pc sample the GCNS, making this sample amenable for use in searching for synchronized signals via the SETI Ellipsoid framework. 
}
\label{fig:derr}
\end{figure}

The accuracy of the expected arrival time for synchronized signals is directly impacted by uncertainties in the distance to the target stars. In Figure \ref{fig:derr} we show the distance uncertainty for GCNS stars as a function of distance. Here we follow \citet{bailer-jones2021}, and use the ``symmetrized distance uncertainty'' from the provided distance probabilities (16\%, 50\%, 84\%), also known as ($r_{lo}$, $r_{med}$, $r_{hi}$), computed as $\sigma_r = (r_{hi} - r_{lo})/2$. The distance shown in Figure \ref{fig:derr}, and used throughout this work, is the 50\% probability ($r_{med}$). As \citet{bailer-jones2021} describe, the distance posterior probabilities depend on the star's distance, color, and variations in the Gaia limiting magnitude. While there is up to 2 dex of range in Figure \ref{fig:derr}, typical (median or mode) distances uncertainties are less than 0.1 lyr for stars out to $\sim$100 lyr, and less than 1 lyr for stars out to 326 lyr (100 pc). From the total GCNS sample, 4\% of stars have symmetrized distance uncertainties less than 0.1 lyr, and 
54\% of stars less than 1 lyr in Gaia EDR3. This provides a large number of nearby targets whose SETI Ellipsoid crossing times (i.e. when we could expect to receive synchronized signals) are accurate to within $\sim$1 year or better.

%%%%%%%%%%%%%%%%%%%%%%%%%%%%%%
\section{Selecting Targets on the \\SN 1987A SETI Ellipsoid }
\label{sec:targets}

With robust distances to stars from Gaia, and a synchronizing event identified (e.g. SN 1987A), selecting targets to monitor using the SETI Ellipsoid requires solving the linear Equation \ref{eqn} based on the elapsed time since the synchronizing event was observed. The Ellipsoid dimensions (i.e. $\mathscr{A}$ and $\mathscr{B}$) grow with time. 
Since the Ellipsoid grows with the speed of light, which is must faster than the relative motions of stars within our Galaxy, we assume that all objects are stationary, and thus $d_1$ and $d_2$ don't change in time for a given star.

%%%%%
\begin{figure}[]
\centering
\includegraphics[width=3.25in]{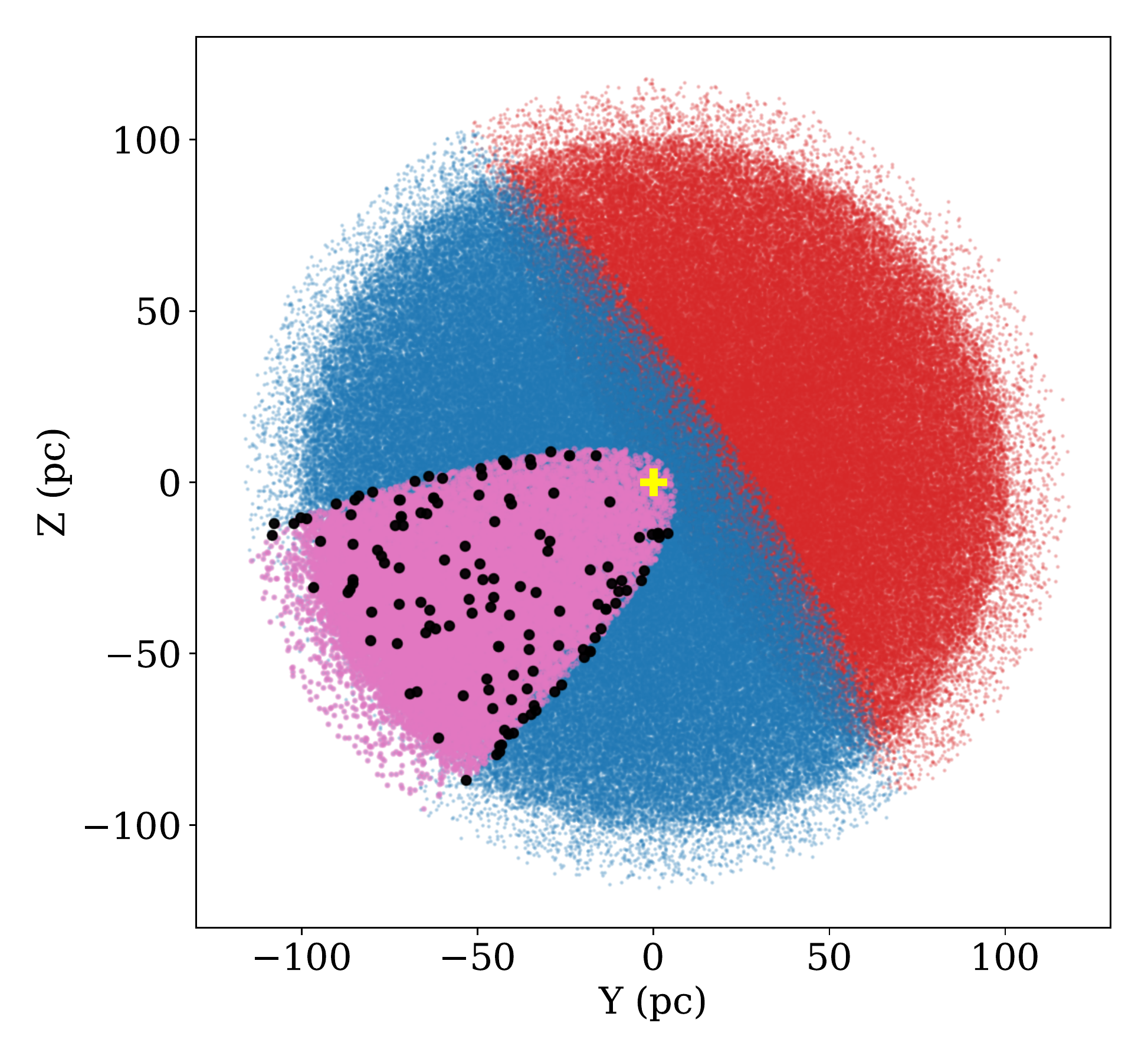}
\caption{
Galactocentric Y,Z location for stars in the GCNS, with points colored by their designation within the SETI Ellipsoid framework for SN 1987A. \xblue stars have seen SN 1987A but we could not yet have received a synchronized transmission (blue points). \xred stars have not observed SN 1987A (red points). There are \xpink stars inside the current SETI ellipsoid for SN 1987A (purple points). At time of writing (mid 2022), \xblack stars from the GCNS are within 0.1 lyr of the SN 1987A Ellipsoid surface. The location of the Sun is also indicated (yellow cross).
}
\label{fig:GCNS}
\end{figure}

For the GCNS, we demonstrate this selection in Figure \ref{fig:GCNS} by finding stars within 0.1 lyr of the present day SN 1987A SETI Ellipsoid. The threshold of 0.1 lyr was chosen to illustrate the near best possible synchronization timing precision that Gaia distances can achieve. The actual value of this Ellipsoid threshold should be chosen with properties of the monitoring campaign in mind, such as observing baseline and cadence.

We find that presently \xblack stars from the GCNS are within 0.1 lyr of the SN 1987A SETI Ellipsoid, i.e. are within $\pm$0.1 years of having possible synchronized signals arrive to us. These targets lay on a cone-like distribution oriented towards the LMC.  From the 331k stars in the GCNS, \xpink stars are inside the Ellipsoid, meaning we can no longer receive synchronized signals from them. \xblue stars have seen SN 1987A but are too far away for us to have received a synchronized transmission yet. \xred GCNS sources have not observed SN 1987A.

A technosignature monitoring campaign of nearby stars starting now should consider prioritizing these \xblack stars. We emphasize however this target list will change steadily over time. At present we find that an average of 734 stars per year within the GCNS intersect the SN 1987A Ellipsoid. While this is a large number of targets to monitor each year, it is well within the capability for many surveys or (semi-)robotic follow-up instruments  \citep[e.g.][]{lsst}.

\begin{figure}[]
\centering
\includegraphics[width=3.25in]{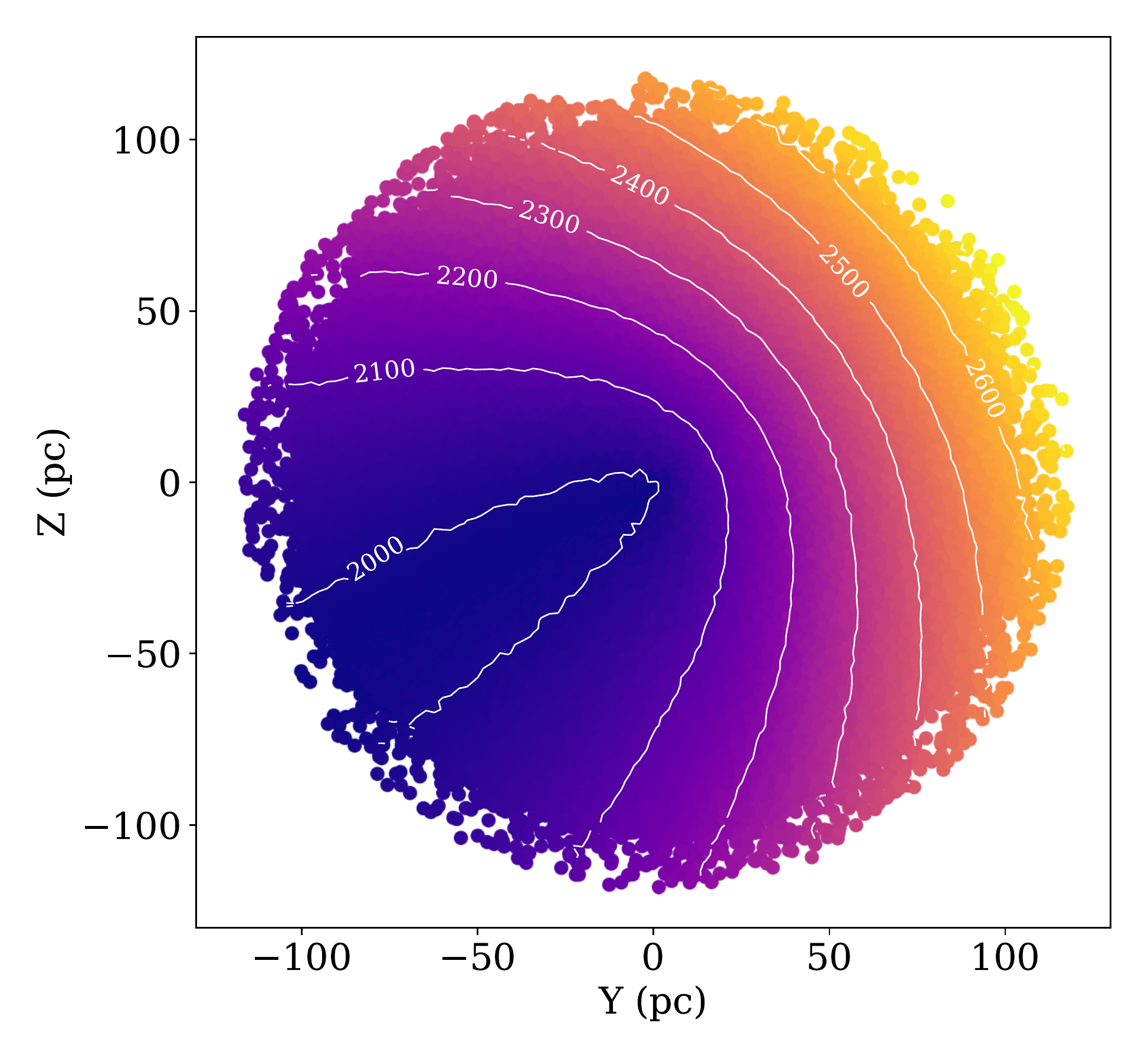}
\caption{
Earth-centered galactocentric Y,Z positions for a slice of 47,484 stars within the GCNS with $|\rm{X}|<10$ pc. Points are colored by the date when each star star intersects the SN 1987A SETI Ellipsoid. Contours show the progression of the SETI Ellipsoid each century.
}
\label{fig:future}
\end{figure}

To further illustrate the changing sample of SETI Ellipsoid targets over time, in Figure \ref{fig:future} we compute the Ellipsoid crossing date (in calendar years) for the entire CGNS sample. For visual clarity we only show a ``slice'' through the GCNS, limiting the plot to the 47,484 stars with galactocentric $|\rm{X}|<10$ pc. The SETI Ellipsoid for SN 1987A continues to widen significantly over the next $\sim$50 years, but will not reach the furthest stars in this sample for several hundred years.

%%%%%%%%%

To explore possible signs of synchronized SN 1987A signals from these \xblack stars in the GCNS, we cross-matched their positions with the 19,355 published variability ``alerts'' from the Gaia Science Alerts archive \citep{hodgkin2021}. 
This alert stream is designed to identify new and rapid changes in brightness by Gaia over its entire observing lifetime. The pipeline for identifying Gaia alerts from the large volume of real-time data is complex, and involves by-eye validation. As such, this database is not a complete reference for significant variability detected with Gaia, but instead is a useful resource when searching for possible dramatic events. None of the \xblack stars in our SETI Ellipsoid crossing sample were found to have an alert issued. The nearest alert cross-match to one of our \xblack stars was over 200 arcsec away.

However, our sample of \xblack stars contain those that are likely crossing the SETI Ellipsoid {\it now}, and so we should not expect they would have been targets for receiving synchronized signals in the past.
Since Gaia has been observing since mid 2014, we expanded our sample to include stars that crossed the SN 1987A Ellipsoid anytime during the Gaia mission. This recovered a total of 5658 sources in the GCNS. Cross-matching these stars to the Gaia alerts catalog we again find no compelling alerts, with the closest being 63 arcsec away. While this alert database does not contain a complete search for variability from these 5658 stars, we can rule out any long timescale (months to years) or dramatic novae-like technosignatures from this sample. As the Gaia mission continues to publish variability alerts, monitoring the stream for the current sample of SETI Ellipsoid crossing sources is a low-effort method for conducting passive SETI. A more comprehensive study of variability for these sources will be possible with future Gaia data releases.

%%%%%%%%%%%%%%%%%%%%%%%%%%%%%%
\section{Galactic Novae as Synchronizing Events}
\label{sec:future}

Thus far we have used the most recent local supernova, SN 1987A as our focus for searching for signal synchronization. However, there are numerous other conspicuous astronomical phenomena that have been suggested for use in developing the SETI Ellipsoid, including Gamma-Ray bursts \citep{corbet1999}, binary neutron star mergers \citep{seto2019}, and historical supernovae \citep{seto2021}. We cannot know what timescales or astrophysical processes would seem ``conspicuous'' to an extraterrestrial civilization with likely a much longer baseline for scientific and technological discovery  \citep[e.g.][]{kipping2020, balbi2021}. Therefore we acknowledge the potential for anthropogenic bias inherent in this choice, and instead focus on which phenomena may be well suited to {\it our} current observing capabilities.

Galactic (or classical) novae were one of the very first targets suggested for use when developing the SETI Ellipsoid framework \citep{makovetskii1977}. These events have characteristically bright optical outbursts, and occur within the Galaxy at a rate of $\sim$50 per year \citep{shafter2017, de2021}. Novae occur predominantly near the Galactic plane, which means distant events are likely obscured by extinction \citep{kawash2021}. This results in the majority of novae events being relatively nearby (within a few kpc), which could help facilitate signal synchronization over smaller regions of the Galaxy.

\begin{figure}[!ht]
\centering
\includegraphics[width=3.25in]{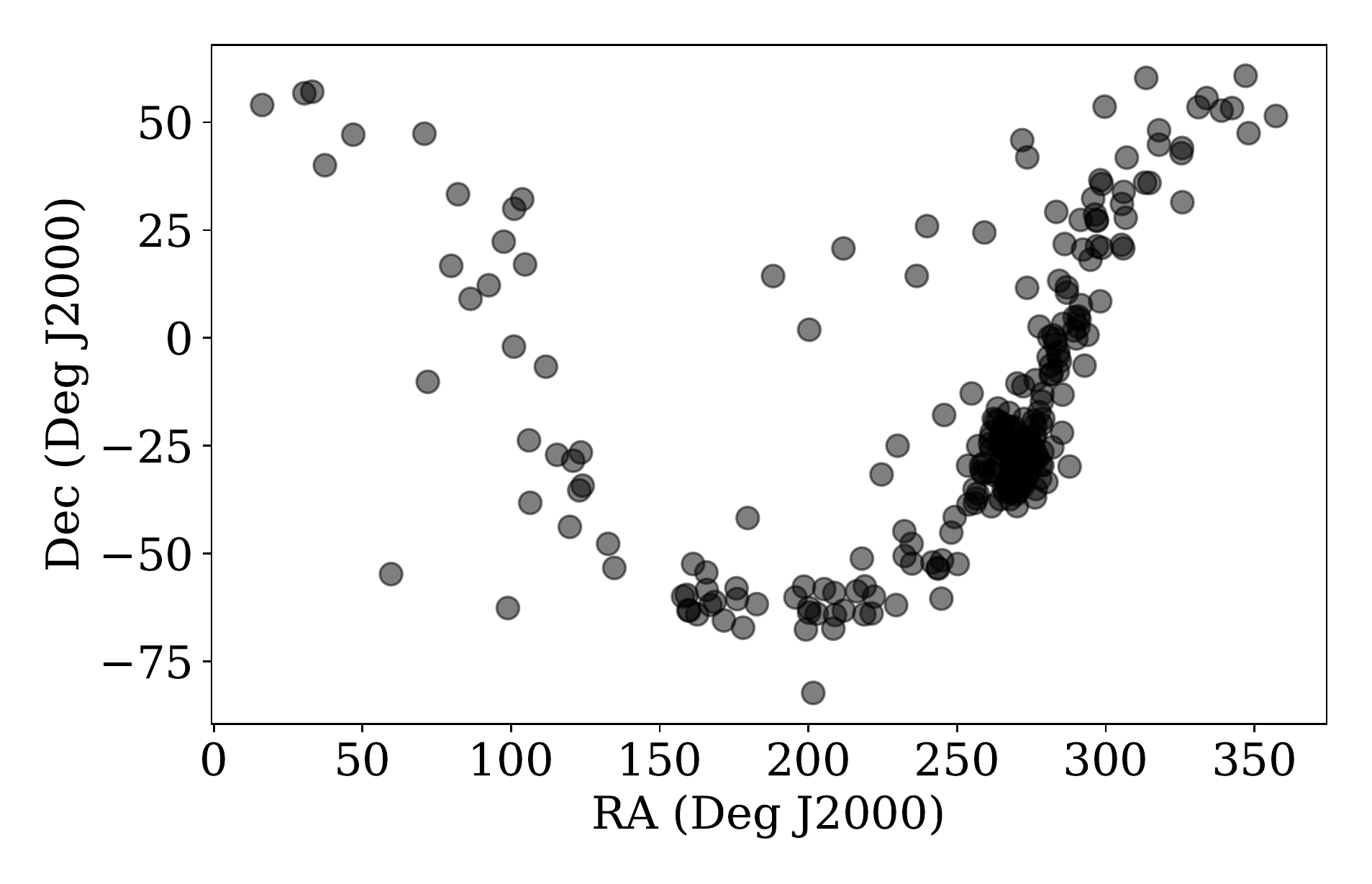}
\includegraphics[width=3.25in]{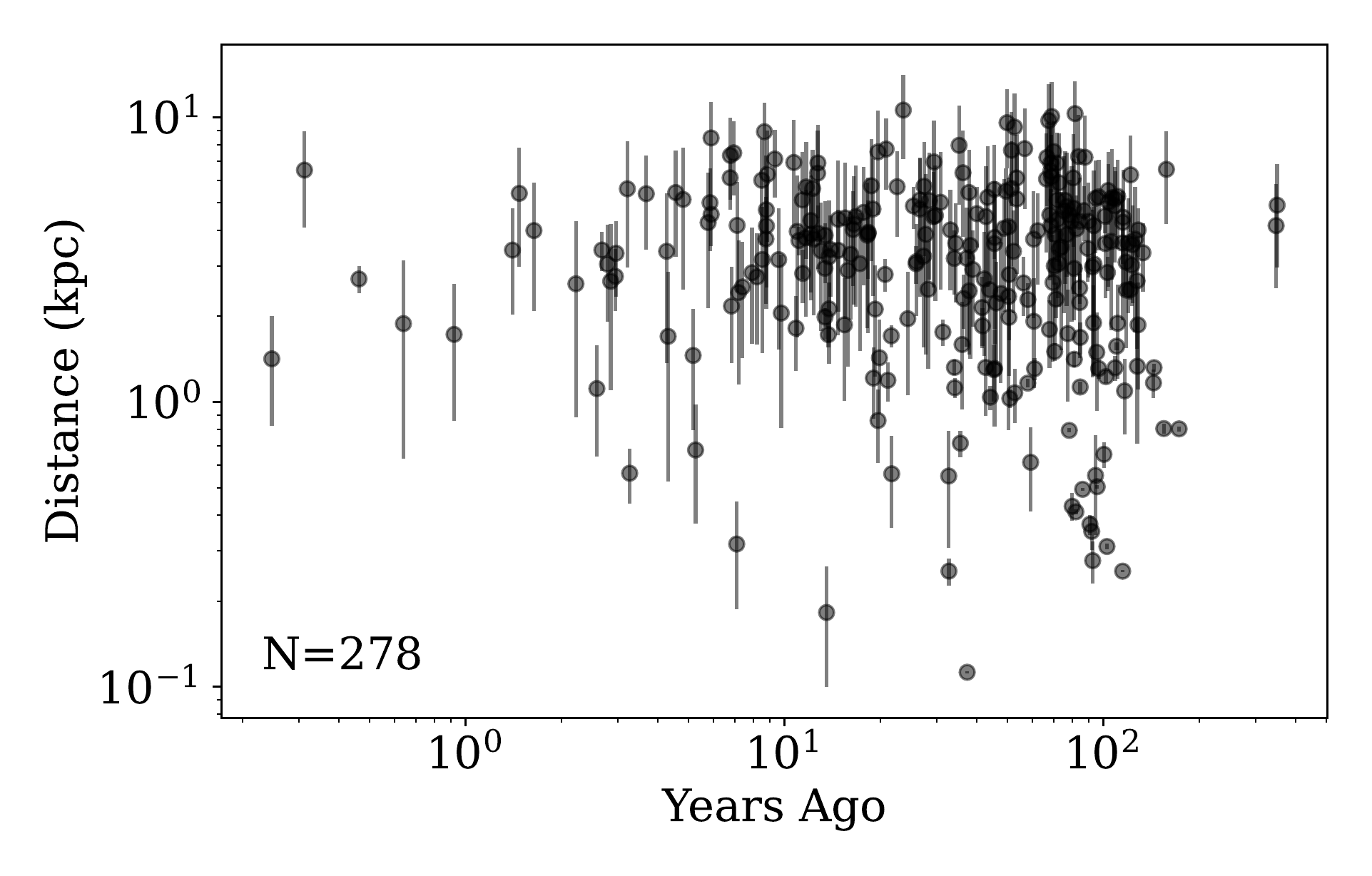}
\caption{
Top: Sky positions for 278 galactic novae with cross-matches from Gaia EDR3 and distances from \citet{bailer-jones2021}. These events are predominantly located within the galactic midplane.
Bottom: Distance versus time since discovery for the 278 novae in our sample. Large uncertainties in distance are the result of faint quiescent luminosities for the novae systems in Gaia, and high extinction.}
\label{fig:novae}
\end{figure}

To explore the utility of classical novae for the SETI Ellipsoid in the Gaia era, we analyzed a large set of events with Gaia EDR3 distances. Our catalog of novae events was drawn from the list maintained by Bill Gray\footnote{https://github.com/Bill-Gray/galnovae}, which includes both historical novae events and regularly updated contributions from ongoing wide field surveys. 
We cross matched this list with Gaia EDR3 and the probabilistic distances from \citet{bailer-jones2021} using a 3 arcsec matching radius, and selected 278 events with robust distances. As shown in Figure \ref{fig:novae}, these events occur primarily near the galactic plane, and have been detected out to distances of $\sim$10 kpc. This sample includes novae that were observed hundreds of years ago, with the oldest being Novae 1670, also known as CK Vul \citep{shara1985}.

\begin{figure}[!ht]
\centering
\includegraphics[width=3.25in]{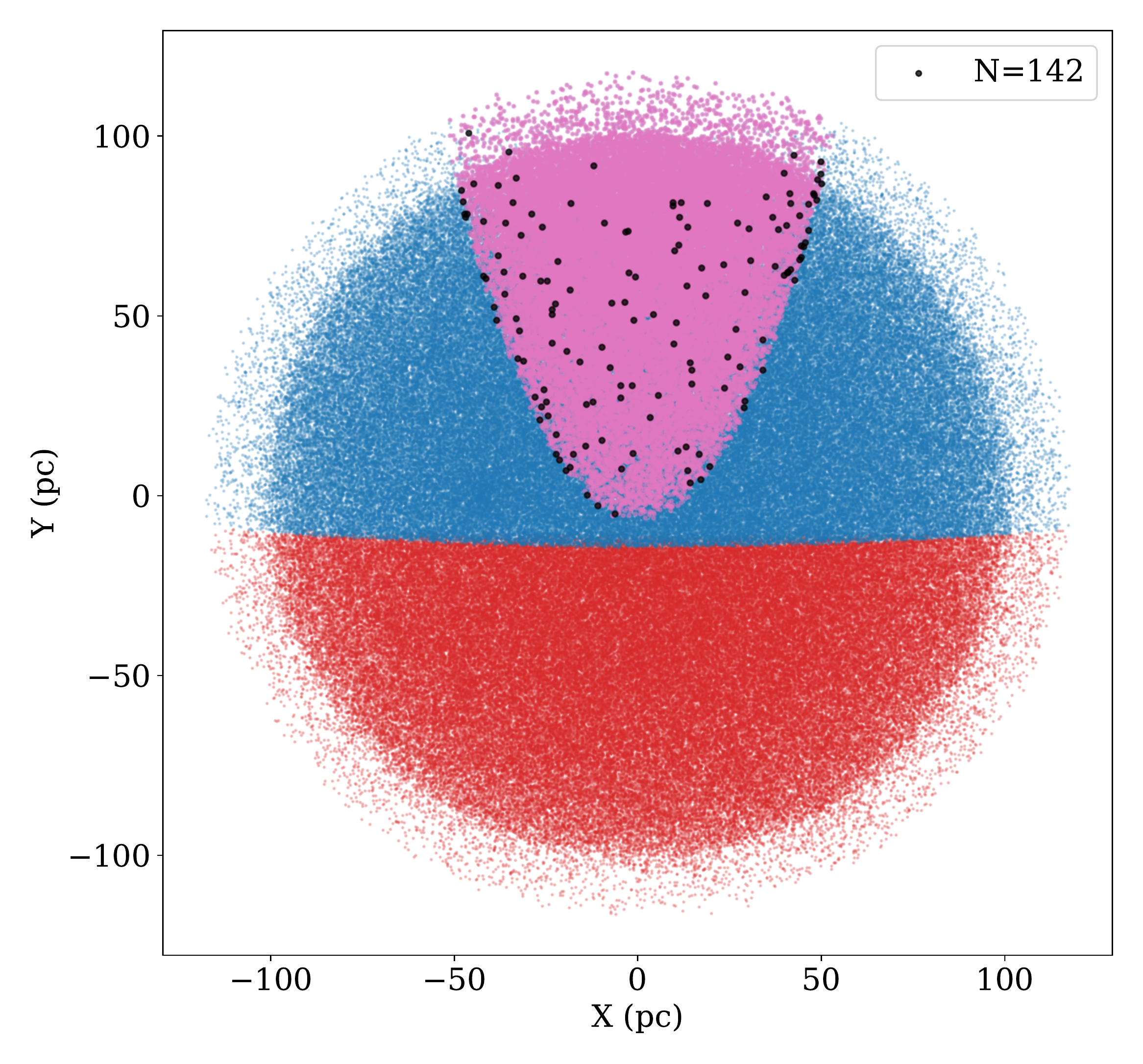}
\caption{
SETI Ellipsoid diagram for Novae Cygnus 1975 in galactocentric X,Y coordinates, centered on the Earth.
Colors for points are the same as in Figure \ref{fig:GCNS}. 142 sources with the GCNS are presently located within 0.1 lyr of the SETI Ellipsoid for N Cyg 1975.
}
\label{fig:cyg75}
\end{figure}

As a representative example from these 278 novae, in Figure \ref{fig:cyg75} we show the current day SETI Ellipsoid for Novae Cygnus 1975. This is the same event \citet{makovetskii1977} originally suggested using for defining the SETI Ellipsoid. While the Ellipsoid has broadened considerably over the past 45 years, and the opportunity for us to broadcast a synchronized signal to nearby stars has passed, N Cyg 1975 remains a viable event for identifying stars to monitor.

These events are fairly frequent, up to $\sim$10 per year in the current data. For each of the 278 novae we computed their current-day SETI Ellipsoid profiles, and selected stars within 0.1 lyr of the Ellipsoid surface. 
Each event had $\sim$140 stars near the Ellipsoid surface, comparable to the point-in-time estimate for the SN 1987A Ellipsoid of \xblack. This is expected, since the yield of sources within a fixed tolerance of the Ellipsoid is directly related to the local stellar density. In total, SETI Ellipsoids from the 278 novae events had 41,180 source intersections identified from the GCNS, coming from 36,135 individual stars (i.e. 5045 stars currently intersect two or more SETI Ellipsoids).

As with the SN 1987A example above, we repeated this exercise for the 278 galactic novae SETI Ellipsoids including any GCNS stars that intersected the Ellipsoid since mid 2014. We again matched the Gaia Science Alerts archive to these Ellipsoid intersections. We only considered alerts within 1 arcsec of a SETI Ellipsoid crossing star, stars that had distance uncertainties less than 1 lyr, and alerts that occurred after the expected Ellipsoid crossing time (allowing for a 0.3 year window {\it before} the Ellipsoid crossing time to account for timing uncertainty).

For 16 of the 278 novae we recovered a Gaia Science Alert of interest. These alerts were all classified in the Gaia Science Alerts archive as coming from nearby flare or cool stars, with the exception of an outburst from the known CV, VW Hyi in early 2022 (Gaia Alert: {\tt Gaia22alg}), which intersected the SETI Ellipsoids for both N Sco 1893 and N Mus 1983 at different times in 2015. VW Hyi is highly variable, with more than a dozen outbursts throughout the light curve provided with the Gaia Alert data. We therefore do not believe any of the events recovered from the Gaia Science Alerts represent a synchronized signal along the SETI Ellipsoid from these 278 galactic novae.

All together, SETI Ellipsoids based on classical novae yield a large number of stars (more than 36k) to monitor at any given time (i.e. within 0.1 lyr of the present day Ellipsoids).
This makes galactic novae likely too numerous to use for in selecting sources for targeted SETI monitoring campaigns. However, as we have demonstrated with the Gaia Science Alerts archive, it is straightforward to mine survey data and real-time alerts for sources that intersect a great many SETI Ellipsoids.

%%%%%%%%%%%%%%%%%%%%%%%%%%%%%%
\section{Discussion}
\label{sec:discuss}

We have presented a review of the SETI Ellipsoid framework for identifying signals that could be synchronized with conspicuous astronomical events. The remarkably precise parallaxes provided by the Gaia mission enable accurate distance estimates for stars within 100 pc of the Sun. These in turn allow us to constrain timing uncertainties for stars intersecting the SETI Ellipsoid. Gaia allows the SETI Ellipsoid technique to finally be useful in selecting targets for robust technosignature searches. The precision mapping of stars from Gaia also enables other methods for constraining technosignature targets based on 2D and 3D positions, e.g. the Earth Transit Zone \citep{heller2016,wells2018}.

Gaia and the SETI Ellipsoid are a powerful basis for enabling technosignature searches with modern wide-field surveys \citep{djorgovski2000, davenport2019a}. 
The Ellipsoid algorithm is straightforward to implement, and we recommend it be used in searching data from many time domain surveys that can be cross matched to Gaia, as well as selecting targets for SETI monitoring programs \citep{isaacson2017}. 
The SETI Ellipsoid framework provides a clear {\it when} and {\it where} to consider in selecting targets or mining large data archives. However, this approach does not specify what type of signal we might expect to be synchronized. There is a clear need for further theory development around the types of technosignatures that can be explored with modern surveys \citep{sheikh2020}.

We have demonstrated the utility of the SETI Ellipsoid in selecting targets over time from the GCNS, and in mining variability information from the Gaia Science Alerts. 
The best timing accuracy for exploring signal synchronization is available for stars with the smallest distance uncertainties. We propose that the 4\% of nearby stars within the GCNS (13,789 sources) with $<$0.1 lyr symmetrized distance uncertainties be a priority for future monitoring when they cross SETI Ellipsoids of interest. These sources have typical distances $<$100 pc, and a median brightness of $G=12.3$ mag.

The nearly volume complete nature of the GCNS means we can place robust limits on the the technosgnature searches we carry out for these stars. However we emphasize the Gaia Science Alerts used here do not contain a complete census of variability from these stars, and instead are illustrative of the type of search that can be done with light curves from forthcoming Gaia data releases (e.g. DR3, expected mid-2022). The continued improvement in parallax precision from future Gaia releases will also expand the sample of stars with acceptably small distance uncertainties that should be monitored. 
In future work, we will present practical explorations using the SETI Ellipsoid framework for target selection from both space-based \citep[TESS;][]{tess} and ground-based \citep[ZTF;][]{bellm2019} surveys.

%%%%%%%%%%%%%%%%%%%%%%%%%%%%%%
\acknowledgments

The authors wish to thank Jill Tarter and Jason Wright for helpful discussions that contributed to the development of this project. 
% \\

JRAD acknowledges support from the DiRAC Institute in the Department of Astronomy at the University of Washington. The DIRAC Institute is supported through generous gifts from the Charles and Lisa Simonyi Fund for Arts and Sciences, and the Washington Research Foundation.

The authors acknowledge support from the Breakthrough Listen initiative. Breakthrough Listen is managed by the Breakthrough Initiatives, sponsored by the Breakthrough Prize Foundation.

SC acknowledges support as the Director of the 
Berkeley SETI Research Center Research Experience for Undergraduates Site, supported by the National Science Foundation under Grant AST 1950897.

S.Z.S. acknowledges that this material is based upon work supported by the National Science Foundation MPS-Ascend Postdoctoral Research Fellowship under Grant No. 2138147.

We acknowledge ESA Gaia, DPAC and the Photometric Science Alerts Team (http://gsaweb.ast.cam.ac.uk/alerts).

\software{
Python, IPython \citep{ipython}, 
NumPy \citep{numpy}, 
Matplotlib \citep{matplotlib}, 
SciPy \citep{scipy}, 
Pandas \citep{pandas}, 
Astropy \citep{astropy}
}

%\bibliography{references.bib}

\end{document}